%% file: main.tex
\documentclass[letterpaper]{article}
\usepackage{aaai}
\usepackage{times}
\usepackage{helvet}
\usepackage{courier}
\usepackage[colorinlistoftodos,textsize=tiny]{todonotes}
\usepackage{multirow}

\usepackage{booktabs}
\usepackage{nicefrac}
\usepackage{amsmath,bm,dsfont,bbm,mathtools}
\usepackage{algorithmic}
\usepackage{amsthm,nccmath,lipsum,cuted}
\usepackage{xcolor}
\usepackage{colortbl}
\usepackage{graphicx}
\usepackage[lined,boxed,commentsnumbered, ruled,linesnumbered]{algorithm2e}
\usepackage{subcaption}
\usepackage{cleveref}

\usepackage{amsfonts}
\usepackage{ragged2e}

\usepackage{graphicx}
\usepackage{verbatim}
\usepackage{supertabular}
\usepackage{multicol}
\usepackage{physics}
\usepackage{comment}
\usepackage{dblfloatfix} 
\usepackage[footnote,draft,danish,silent,nomargin]{fixme}

\begin{document}
%

\author{Adarsh Prasad Behera\textsuperscript{\rm 1}, Roberto Morabito\textsuperscript{\rm 2}, Joerg Widmer\textsuperscript{\rm 1}, Jaya Prakash Champati\textsuperscript{\rm 1}\\
\textsuperscript{\rm 1}IMDEA Networks Institute, Madrid, Spain\\
\textsuperscript{\rm 2}University of Helsinki, Helsinki, Finland
}

\title{Improved Decision Module Selection for Hierarchical Inference in Resource-Constrained Edge Devices\thanks{This research is funded by the European Union through MSCA-PF project “DIME: Distributed Inference for Energy-efficient Monitoring at the Network Edge” under Grant Agreement No. 101062011.\\
This work was presented at Deployable AI Workshop (DAI) at AAAI-2024.}}

\maketitle
\begin{abstract}
\begin{quote}
The Hierarchical Inference (HI) paradigm employs a tiered processing: the inference from simple data samples are accepted at the end device, while complex data samples are offloaded to the central servers. HI has recently emerged as an effective method for balancing inference accuracy, data processing, transmission throughput, and offloading cost. This approach proves particularly efficient in scenarios involving resource-constrained edge devices, such as IoT sensors and micro controller units (MCUs), tasked with executing tinyML inference. Notably, it outperforms strategies such as local inference execution, inference offloading to edge servers or cloud facilities, and split inference (i.e., inference execution distributed between two endpoints). Building upon the HI paradigm, this work explores different techniques aimed at further optimizing inference task execution. We propose and discuss three distinct HI approaches and evaluate their utility for image classification. 
\end{quote}
\end{abstract}


\input{introduction.tex}

\input{methodologies.tex}

\input{findings.tex}

\input{conclusion.tex}

\bibliographystyle{aaai}


\end{document}

%% file: introduction.tex
\section{Introduction}

\begin{figure}[t]
  \centering
  \subcaptionbox{TinyML}[.47\linewidth][c]{%
    \includegraphics[width=0.22\textwidth]{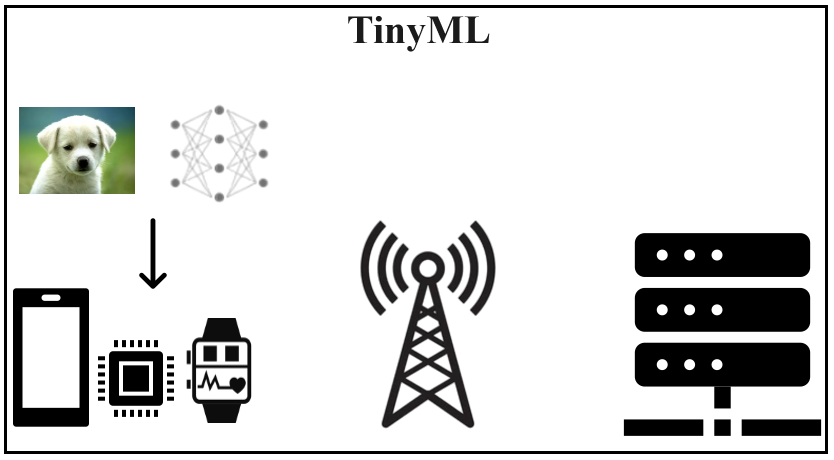}}\quad
  \subcaptionbox{Full offload}[.47\linewidth][c]{%
    \includegraphics[width=0.22\textwidth]{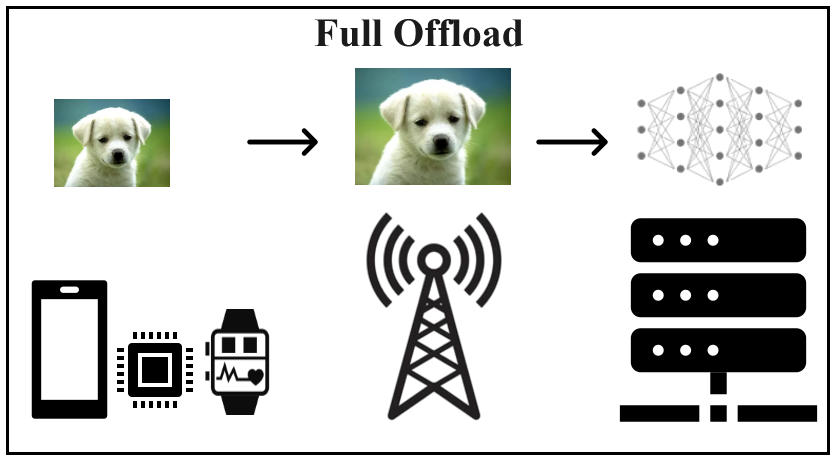}}\quad \\
    \subcaptionbox{Inference load balancing}[.47\linewidth][c]{%
    \includegraphics[width=0.22\textwidth]{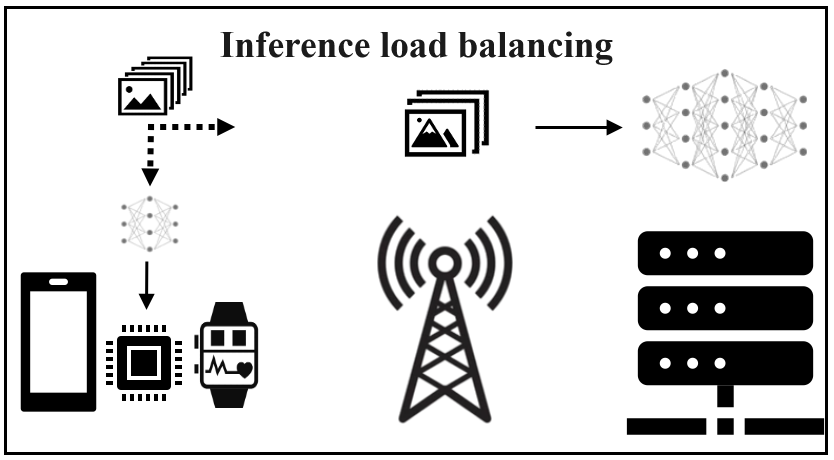}}\quad
  \subcaptionbox{DNN partitioning}[.47\linewidth][c]{%
    \includegraphics[width=0.22\textwidth]{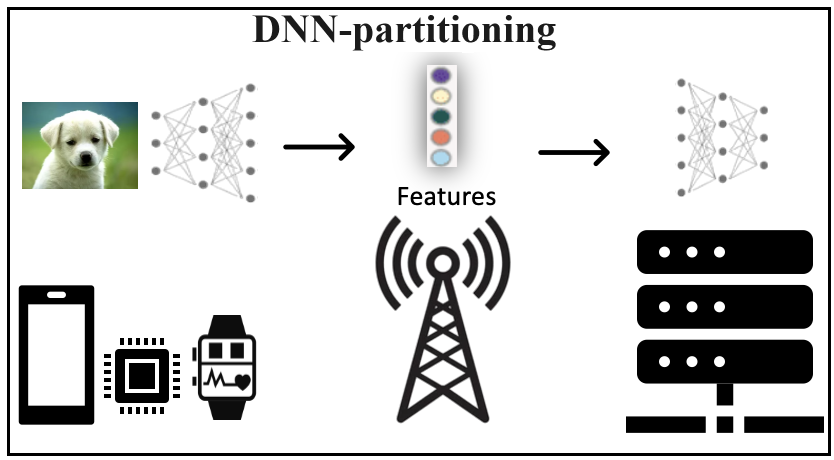}}\quad \\
    \caption{Various approaches for DL inference at the edge.}
    \label{related_tech}
\end{figure}

Machine Learning (ML) inference, the process of utilizing trained ML models to make predictions on new data, is becoming increasingly important due to the rise in applications requiring this capability. This has triggered a significant expansion in research and development activities, particularly in the field of distributed ML inference at the network edge \cite{chen2019deep}. However, performing ML inference on resource-constrained End Devices (EDs) presents unique challenges due to their limited computational capabilities and energy resources. To overcome these constraints, a variety of strategies have been proposed as depicted in Fig. \ref{related_tech}. These encompass the use of tinyML, a branch of ML that focuses on designing lightweight ML models that can operate on EDs with minimal resources \cite{banbury2021mlperf}. A common approach in edge computing is full offload that involves transferring entire computational tasks from ED to powerful remote resources typically Edge Servers (ESs) or cloud platforms for efficient, resource-intensive processing and analysis. Another strategy is edge intelligence inference load balancing, where computationally intensive tasks are transferred from the ED to ESs or cloud platforms \cite{fresa2022offloading}. Deep Neural Network (DNN) partitioning is yet another technique that divides the inference execution of large-scale DNNs between an ED and an ES
\cite{huang2020clio}. While these approaches can mitigate the challenges of executing ML inference on EDs, they can introduce other issues such as latency and increased communication costs.

\begin{figure} [t!] 
    \begin{center}
  \includegraphics[width=8cm]{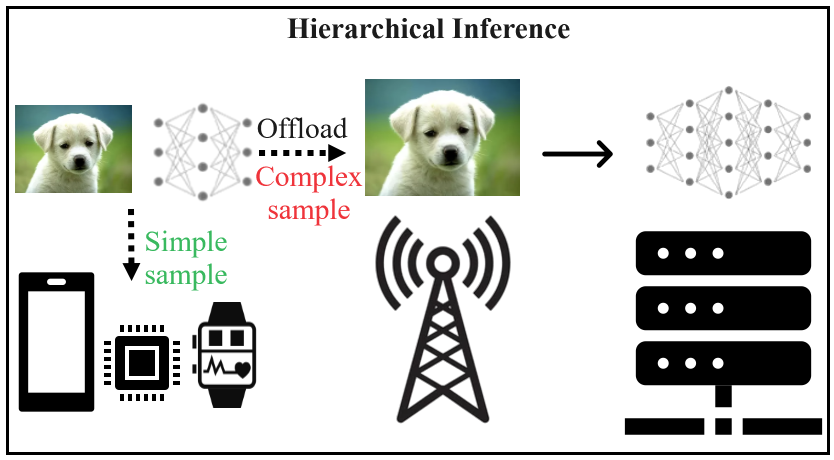}
    \end{center}
    \caption{HI framework for DL inference at network edge.} \label{fig:HI-system}
\end{figure}

In this context, Hierarchical Inference (HI) emerges as a novel framework for performing distributed Deep Learning (DL) inference at the edge \cite{moothedath2023online,al2023case}. In a HI system, an ED is equipped with a tinyML model (referred to as Small ML or S-ML) , and it can also offload the inference tasks to a state-of-the-art large-size ML model (L-ML) residing on an ES or cloud. A data sample is deemed \textit{simple data sample} if S-ML inference suffices. Differently, a \textit{complex data sample} necessitates L-ML inference. The underlying idea of HI is that only complex data samples should be offloaded to the ES or cloud. The HI framework is shown in Fig. ~\ref{fig:HI-system}. The decision module of HI utilizes the S-ML inference output and associated metadata, such as application QoS requirements and the corresponding L-ML model, to determine whether a sample is complex or simple, and subsequently, whether to offload it or not. 
Unlike traditional tinyML research, HI introduces the flexibility of inference offloading. It scrutinizes the S-ML inference first before deciding whether offloading is necessary, in contrast to existing inference load balancing algorithms \cite{fresa2022offloading,xiao2022edge}. HI offers a balanced compromise between tinyML and DNN partitioning methodologies. With HI, it is possible to take advantage of the resource efficiency of S-MLs on EDs while maintaining the option to utilize more robust, state-of-the-art L-MLs. However, HI can only be implemented effectively if the S-ML fulfills certain conditions: \emph{(i)}  the size of S-ML should be small enough for seamless execution on EDs, \emph{(ii)} the energy required for S-ML inference should be lower than that needed for transmitting a data sample, and \emph{(iii)} the accuracy of S-ML should be such that the proportion of simple data samples exceeds that of complex data samples.

Another critical challenge in the implementation of HI is the identification of complex data samples for offloading. To this end, previous research on HI \cite{moothedath2023online,al2023case} used a threshold on the highest softmax value from the final layer of S-ML embedded in the ED.  However, they had to offload more than $35\%$ of total data samples for CIFAR-10 image classification, which increased the cost per image significantly. 
While HI does result in an overall accuracy improvement, its full potential is yet to be realized using more effective methods that go beyond the use of threshold on the softmax value for differentiating complex and simple data samples.

In this work, false negatives (FN) are simple data samples that are offloaded and false positives (FP) are complex data samples that are not offloaded by HI.
Both FNs and FPs contribute to the cost (energy/delay) overhead and must be minimized to achieve optimal system performance. To this end, we propose three novel HI approaches for differentiating complex and simple data samples and evaluate their performance using a state-of-the-art tinyML model  
for CIFAR-10 image classification.



%% file: methodologies.tex
\section{Proposed Methodologies and Implementation}

In this section, we explore several distinct approaches aimed at overcoming these limitations and further increasing the effectiveness and utility of the HI paradigm. 
\begin{enumerate}
    \item Calibration of tinyML model with fixed threshold.
    \item Use of classical ML classifiers after tinyML inference.
    \item Use of classical ML classifiers before tinyML inference.
\end{enumerate}
Although another DNN could be used as the decision module for the offload paradigm, we use simple ML classifiers like Logistic Regression (LR), Support Vector Machine (SVM) or Random Forest (RF) due to the limited memory constraints of MCUs used as EDs in our resource constrained set up. DNNs are complex algorithms and require much more resources (memory and inference energy) as compared to these classical ML classifiers.

In this work, we use a customized ResNetv1 developed by the MLPerf group \cite{banbury2021mlperf} as our S-ML model on the ED. The model's size is $311$ KB, with an impressive benchmark accuracy of $85\%$, making it the ideal choice for our experiments. Moreover, we have considered a pre-trained Vision Transformer Huge model (ViT-H/14) \cite{dosovitskiyimage} as our L-ML model present on the cloud or ES. It has a reported test accuracy of $99.5\%$, which is the highest reported test accuracy on the CIFAR-10 dataset till date.

\subsection{Calibration of TinyML with Fixed threshold}
Confidence calibration is a key component of probabilistic ML models. It reflects of the congruence between model predictions and the actual probabilities or levels of confidence \cite{zhang2023survey}. ML models frequently exhibit overconfidence \cite{guo2017calibration}, which results in incorrect predictions. Overconfidence occurs when a model assigns high probabilities to incorrect or uncertain predictions. 
Calibration can be be mathematically defined as:
\begin{equation}
    \mathbb{P}\left(\hat{y}=y|\hat{p}=p\right)=p , \hspace{2mm}\forall p \in [0,1]
\end{equation}
where $y$ and $\hat{y}$ represent ground truth and predicted class of any data sample respectively and $\hat{p}$ represents the confidence of the neural network (NN) for that particular prediction.

\begin{figure}[t]
  \centering
  \subcaptionbox{Before Calibration}[.47\linewidth][c]{%
    \includegraphics[width=0.21\textwidth]{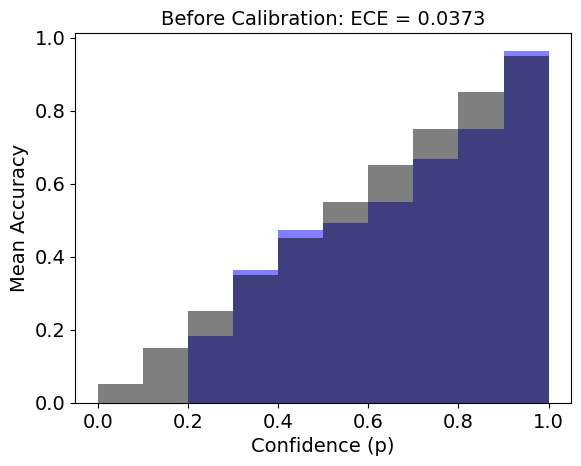}}\quad
  \subcaptionbox{After Calibration}[.47\linewidth][c]{%
    \includegraphics[width=0.21\textwidth]{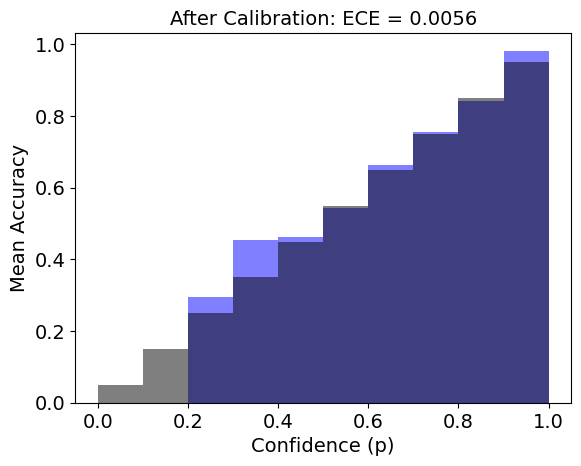}}\quad
    \caption{Mean Accuracy vs Confidence in each bin with respective ECE before and after calibration.}
    \label{Calibration}
\end{figure}

\begin{figure}[t]
  \centering
  \subcaptionbox{Accuracy vs threshold ($\theta$)}[.47\linewidth][c]{%
    \includegraphics[width=0.22\textwidth]{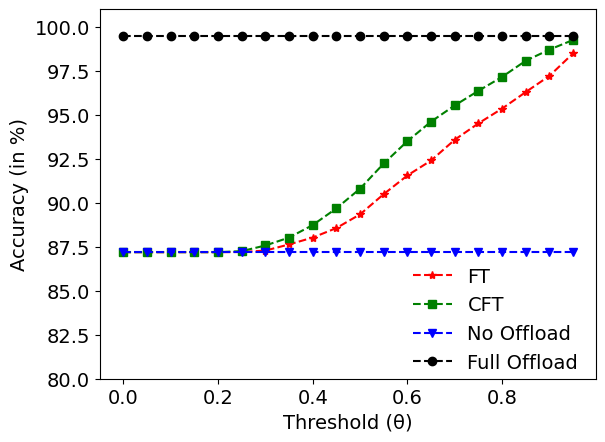}}\quad
  \subcaptionbox{Selection of $\theta^*$ for $\beta=0.5$}[.47\linewidth][c]{%
    \includegraphics[width=0.22\textwidth]{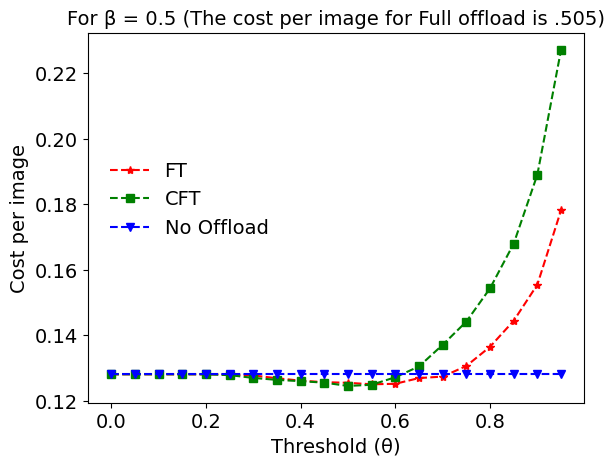}}\quad
    \caption{Optimal threshold ($\theta^*$) selection.}
    \label{parameter_selection}
\end{figure}


In this approach, we calibrate the tinyML model using a well-known post-processing calibration method "Temperature Scaling" \cite{guo2017calibration} to make the probability estimates more reliable and decide an optimal threshold $\theta^*$ for offloading decisions with minimal FPs and FNs. 
To estimate the expected accuracy from finite samples, we group predictions into $10$ interval bins (each of size $0.1$). Fig. \ref{Calibration} shows the tinyML model's mean accuracy relative to each bin's confidence. We observe a reduction in the Expected Calibration Error (ECE) to $0.0056$ following calibration, compared to $0.0373$ prior to calibration. As discussed earlier, the energy consumption per tinyML inference, or the associated cost (denoted as $\alpha$), is considered negligible compared to the energy consumption associated with offloading each sample to the ES or cloud (represented by $\beta$, where $\beta = 0.5$ in our case) and the cost of an incorrect inference (denoted as $\gamma$, for which we considered $\gamma = 1$). Fig. \ref{parameter_selection} shows the optimal threshold ($\theta^*$) selection, based on the cost per image for a fixed value of $\beta = 0.5$. It can be observed in Fig. \ref{parameter_selection}(b) that the cost per image (CPI) is lowest at $\theta = 0.5$ and $\theta = 0.55$ respectively for calibrated and non-calibrated model.

\subsection{Use of Classifiers after TinyML inference}

\begin{figure} [t!] 
    \begin{center}
  \includegraphics[width=7cm]{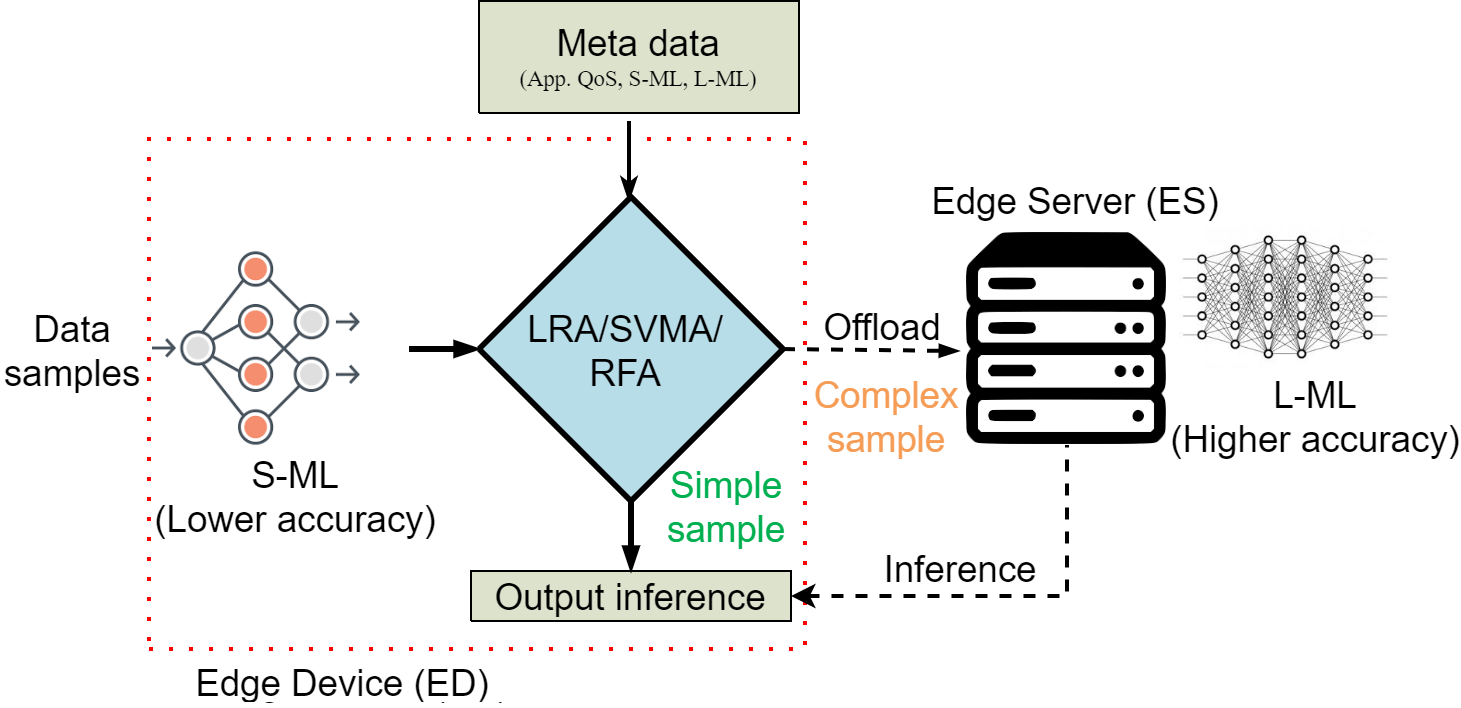}
    \end{center}
    \caption{HI framework with classifiers after the tinyML model on the edge device.} \label{fig:MLCA}
\end{figure}

In this approach, all the images are inferenced at the ED and the probability estimates obtained from the softmax layer of tinyML will be learned using different conventional ML classifiers like LR, SVM and RF on the edge device, and the offloading decision will be made based on that, as shown in Fig. \ref{fig:MLCA}. At first, the tinyML is trained on $50000$ training data and tested on the rest $10000$ images. The complex data samples are identified in these $10000$ images and correspondingly assigned a class (e.g. class $1$) and the rest of the data samples are assigned another class (e.g. class $0$). There is a significant difference in the number of data samples in each of these class which causes class imbalance. Down-sampling is performed on simple data samples to balance the classes. These images are further divided into train and test set in a $80:20$ ratio. The classifiers are trained with the training data and the corresponding weights are saved.



\subsection{Use of Classifiers before TinyML inference}

\begin{figure} [t!] 
    \begin{center}
  \includegraphics[width=6cm]{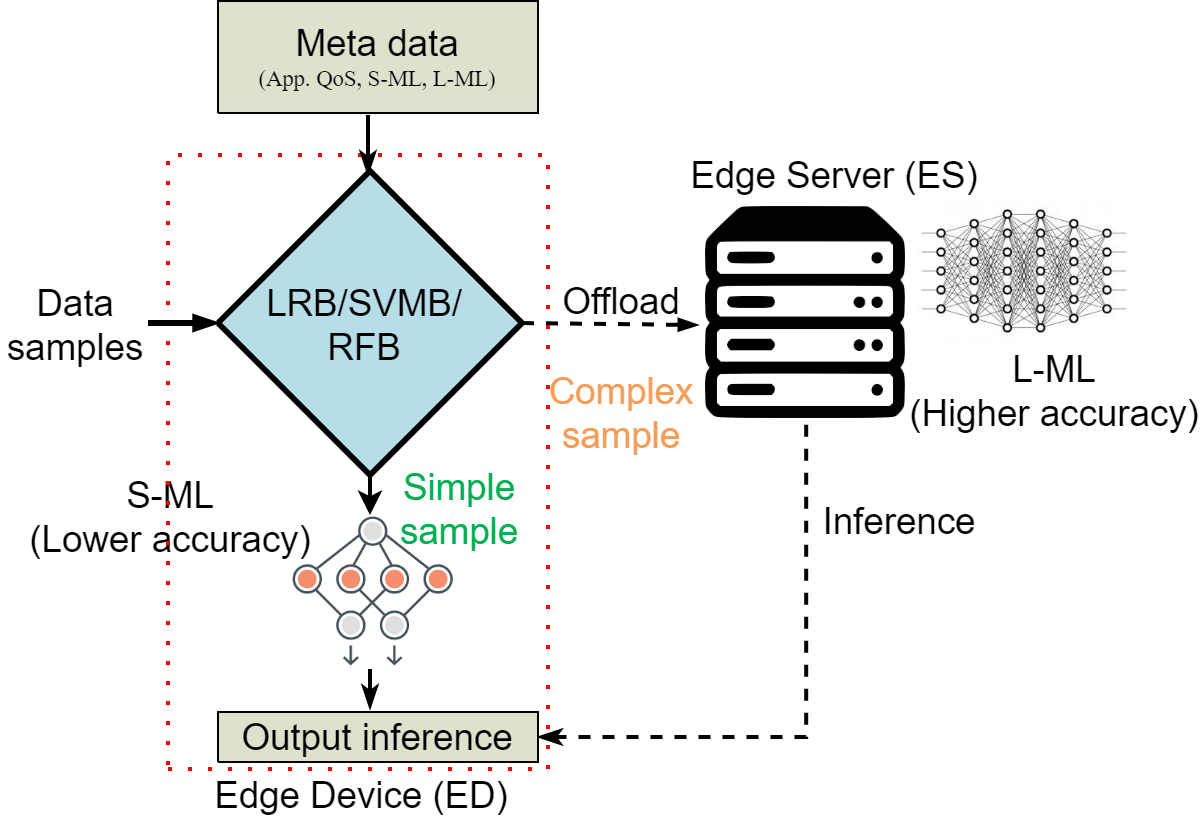}
    \end{center}
    \caption{HI framework with classifiers before the tinyML model on the edge device.} \label{fig:MLCB}
\end{figure}

In scenarios where $\alpha$ is significant in comparison to $\beta$, it could be beneficial to identify complex samples in advance and offload them to the ES or cloud. To do this, we train the classifier models to detect  complex samples before they are assigned to the tinyML for inference, and these samples are directly offloaded to the ES or cloud without being processed on the ED, as can be observed in Fig. \ref{fig:MLCB}. In this case, the classifiers are trained directly on the images to execute a binary classification task, i.e., to predict whether data samples are complex or simple for the given tinyML. 

%% file: findings.tex
\section{Results}

\begin{figure*}[ht]
  \centering
  \subcaptionbox{Cost/image for HI classifiers after tinyML.}[.32\linewidth][c]{%
    \includegraphics[width=0.28\textwidth]{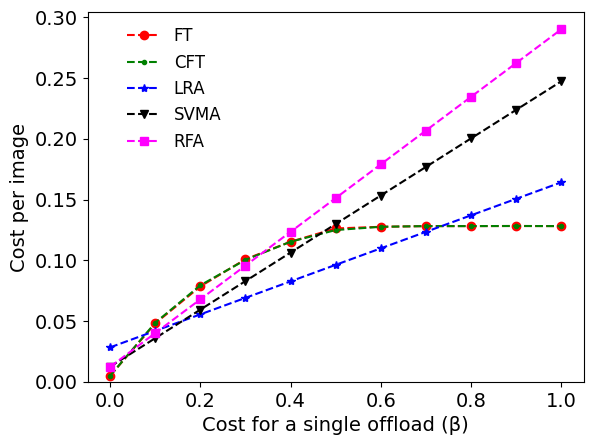}}\quad
  \subcaptionbox{Accuracy under HI classifiers after tinyML}[.32\linewidth][c]{%
    \includegraphics[width=0.28\textwidth]{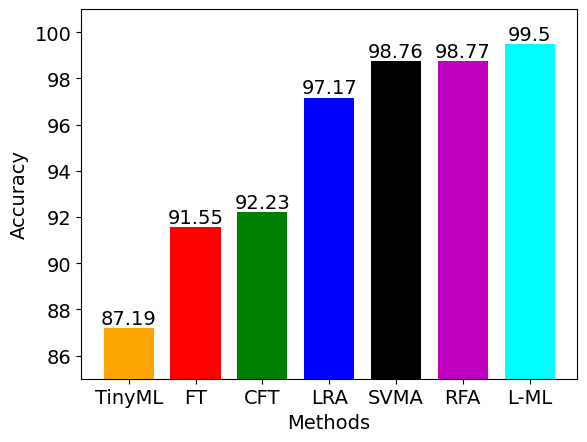}}\quad
  \subcaptionbox{$F1$-score under HI classifiers after tinyML}[.32\linewidth][c]{%
    \includegraphics[width=0.28\textwidth]{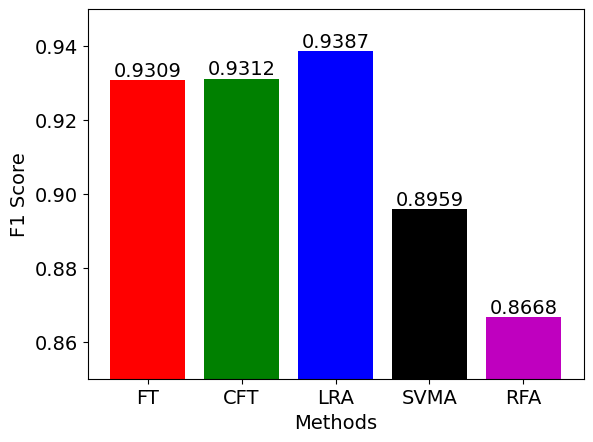}}\quad
    \caption{Comparative Performance Analysis of different approaches when $\alpha\to 0$.}
    \label{results}
\end{figure*}

\begin{figure}[t]
  \centering
  \subcaptionbox{Cost/image for HI classifiers before tinyML and LRA ($\beta=0.5$)}[.47\linewidth][c]{%
    \includegraphics[width=0.22\textwidth]{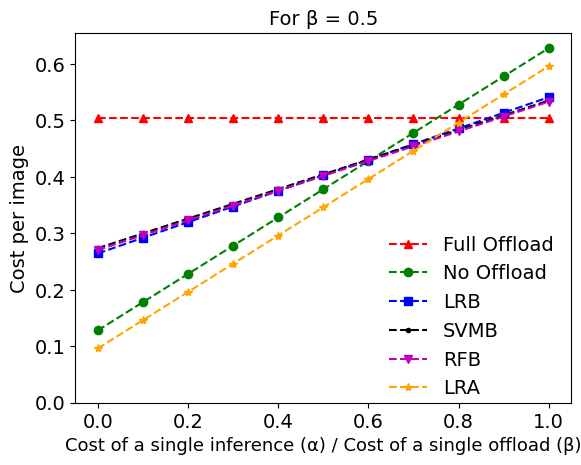}}\quad
  \subcaptionbox{Accuracy under HI classifiers before tinyML and LRA}[.47\linewidth][c]{%
    \includegraphics[width=0.22\textwidth]{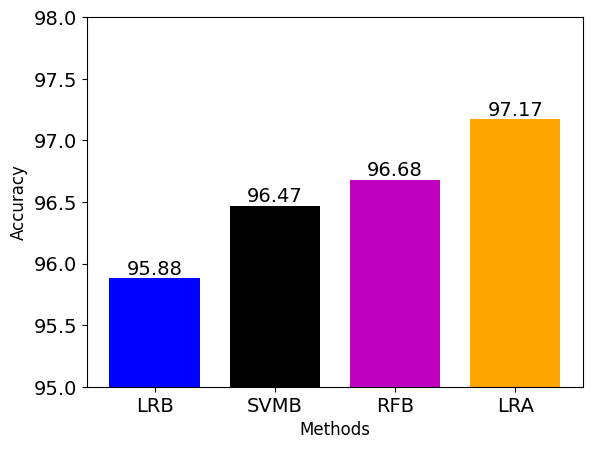}}\quad
    \caption{Comparative Performance Analysis of ML algorithms before TinyML.}
    \label{results2}
\end{figure}

In Fig. \ref{results}, we present the comparative performance analysis of the first two approaches, calibration with fixed threshold (CFT) and ML classifiers after tinyML (LRA, SVMA and RFA) with respect to fixed threshold method (FT) proposed in the previous work \cite{al2023case}. For this case, we assume $\alpha$ is negligible with respect to $\beta$, hence set $\alpha = 0$. It can be observed from Fig. \ref{results}(a), as $\beta$ increases, the cost per image (CPI) increases initially for all the techniques. However, after $\beta$ reached $0.5$, the CPI for FT and CFT stagnated while for the ML classifiers CPI increases linearly with $\beta$. It can also be observed that for the majority of the region ($0.2\leq\beta\leq0.7$) the CPI for LRA is the lowest. Furthermore, in Fig. \ref{results}(c) it can be noticed LRA outperformed all other techniques in terms of  $F1$-score. Although the overall accuracies of SVMA and RFA are a little higher than that of LRA as seen in Fig. \ref{results}(b), LRA outperforms its counterparts due to a lower number of offloads while also restricting the number of FP and FN.

In Fig. \ref{results2}, we consider $\alpha$ is non-negligible relative to $\beta$ and present the comparative performance analysis of ML classifiers (LRB, SVMB and RFB) before tinyML. Fig. \ref{results2}(a) shows CPI for varying $\alpha/\beta$ for a constant $\beta$ value of $0.5$. It is worth noting that as previously discussed one important assumption for HI is the energy required for tinyML inference should be lower than that needed for transmitting a data sample or $\alpha\leq\beta, \forall \beta$. It can be observed in Fig. \ref{results2}(a) that the CPI for each technique increases linearly with an increase in $\alpha/\beta$ value. All three classifiers LRB, SVMB, and RFB perform almost identically. Interestingly, compared to the classifiers before tinyML, CPI for LRA is the lowest till $\alpha/\beta$ reaches $0.7$. There is a small margin $0.7<\alpha/\beta<0.9$, where classifiers before tinyML techniques outperform their counterparts. As $\alpha$ approaches $\beta$ (for $\alpha/\beta\geq0.9$), CPI for full offload became the lowest, suggesting it is better to offload all the data samples rather than doing any inference at the ED. The overall system accuracy for each of the classifiers before tinyML with comparison to logistic regression after tinyML is shown in Fig. \ref{results2}(b).


%% file: conclusion.tex
\section{Conclusion and Future works}
In this work, we proposed three decision module selection strategies for HI that can be implemented to further optimize the offloading performance. Simulation experiments showed that, more often than not, using logistic regression after tinyML (LRA) as the decision module that produced the most favourable outcome in HI, considering cost per image and $F1$-score. In future work, we plan to embed these models into MCUs to measure each technique's energy requirements and validate the overall accuracy achieved in system implementations.